# Simulation study of light transport in laser-processed LYSO:Ce detectors with single-side readout

**L Bläckberg**[1,2], **G El Fakhri**[1] **and H Sabet**[1]

[1] Gordon Center for Medical Imaging, Department of Radiology, Massachusetts General Hospital & Harvard Medical School, Boston, MA, USA

[2] Department of Physics and Astronomy, Uppsala University, Uppsala, Sweden

E-mail: LBlackberg@MGH.Harvard.edu, HSabet@MGH.Harvard.edu

**Abstract.** A tightly focused pulsed laser beam can locally modify the crystal structure inside the bulk of a scintillator. The result is incorporation of so-called optical barriers with a refractive index different from that of the crystal bulk, that can be used to redirect the scintillation light and control the light spread in the detector. We here systematically study the scintillation light transport in detectors fabricated using the Laser Induced Optical Barrier technique, and objectively compare their potential performance characteristics with those of the two mainstream detector types: monolithic and mechanically pixelated arrays. Among countless optical barrier patterns, we explore barriers arranged in a pixel-like pattern extending all-the-way or half-way through a 20 mm thick LYSO:Ce crystal. We analyze the performance of the detectors coupled to MPPC arrays, in terms of light response functions, position histograms, line profiles, and light collection efficiency. Our results show that laser-processed detectors with both barrier patterns constitute a new detector category with a behavior between that of the two standard detector types. Results show that when the barrier-crystal interface is smooth, no DOI information can be obtained regardless of barrier refractive index. However, with a rough barrier-crystal interface we can extract multiple levels of DOI. Lower barrier refractive index results in larger light confinement, leading to better transverse resolution. Furthermore we see that the laser-processed crystals have the potential to increase the light collection efficiency, which could lead to improved energy resolution and potentially better timing resolution due to higher signals. For a laser-processed detector with smooth barrier-crystal interfaces the light collection efficiency is simulated to >44%, and for rough interfaces >73%. The corresponding numbers for a monolithic crystal is 39% with polished surfaces, and 71% with rough surfaces, and for a mechanically pixelated array 33% with polished pixel surfaces and 51% with rough surfaces.

## 1. Introduction

With the recent development of laser induced optical barriers (LIOB), or sub-surface laser engraving (SSLE) techniques, we now have a new category of scintillation detectors whose features fall in between those of the monolithic and the mechanically pixelated scintillators. The LIOB technique has found its way in a variety of medical imaging applications including Positron Emission Tomography (PET) (Moriya et al 2010, Sabet et al 2012a, Hunter et al 2015, Sabet et al 2016a, Uchida et al 2016, Bläckberg et al 2016b), Single Photon Emission Computed Tomography (SPECT) (Sabet et al 2016b), and Computed Tomography (CT) (Bläckberg et al 2016a). In LIOB a high power pulsed laser is tightly focused inside the bulk of a scintillator crystal, causing a local modification of the crystal structure. The modification is manifested by a change in refractive index (RI) of the material, where the modified region will have a refractive index lower than that of the unmodified crystal. The lowest achievable RI value is 1.0, corresponding to void formation in the crystal. The modifications are here referred to as optical barriers (OB), and their size, shape, and refractive index depend on the crystal material as well as the laser parameters (i.e. pulse energy, duration and repetition rate) and the delivery optics that are used during processing. Due to the index mismatch with the crystal bulk the barriers may redirect the scintillation light inside the crystal, and as a result they can be used to control and manipulate the light spread in the detector. Many closely packed optical barriers can form a wall acting similarly to the reflectors inserted in a mechanically pixelated array. The reflectivity of the created wall will depend on the characteristics of individual barriers as well as on how densely these are packed.

While the LIOB technique has been reportedly used for medical imaging applications, there is not much detail with regard to scintillation light transport in laser-processed crystals and how they compare with the two main detector types: monolithic crystals and mechanically pixelated arrays. Furthermore, the nature of the interactions between the laser light and the crystal structure causing the modifications have been explored, but not investigated systematically, and neither have the physical properties of the resulting optical barriers. In this paper, we report on light transport studies of Cerium-doped lutetium yttrium orthosilicate (LYSO:Ce) crystals, which is the mainstay scintillator material for PET. The ultimate goal of this work is the fabrication of a high-sensitivity and high spatial resolution

LYSO:Ce detector with depth of interaction (DOI) information and single-side readout, which is demanded for high resolution small animal PET imaging. The small gantry size in these systems, compared to whole body PET, increases the probability of oblique angles of incidence of the gamma rays on the detector face, which will increase the severity of the parallax error caused by mapping all events to the center of each crystal. There are a number of different approaches, (Peng and Levin 2010, Ito *et al* 2011), that are proposed to obtain DOI information to compensate for the parallax error, including double side readout, phoswich detectors, statistical positioning algorithms for monolithic crystals (Miyaoka *et al* 2008, Ling *et al* 2007b), and analysis of the signal rise time (Wiener *et al* 2013). Work has also been done with more complex reflector arrangements employed to encode the light spread in the detector block as a function of DOI (Ito *et al* 2010). We are exploring the potential of extracting DOI information from the light response functions of scintillators containing optical barriers. Similar work has previously been done for monolithic crystals (Lerche *et al* 2005), as well as for pixelated detectors with depth dependent light sharing (Yang *et al* 2009). Unlike many other complex detector arrangements that are proposed, our approach has the potential of providing a detector with DOI capability and high transversal resolution in a cost-effective manner, without adding to the system complexity.

We propose to use the flexibility of the LIOB technique to create a pattern of optical barriers inside the scintillator volume, in order to spread the scintillation light over the photodetector face in such way that both transverse and DOI resolution can be achieved. The fact that the optical barriers can be placed virtually anywhere, in any pattern, inside the crystal, and even in the entrance windows of the photodetector elements and the light guide, is the key area that differentiates the laser-processed detectors from the standard monolithic and mechanically pixelated ones.

A scintillator detector fabricated using LIOB falls between the two extremes of a monolithic detector with no incorporated structures, and a mechanically pixelated array that relies on near complete optical isolation between pixels, as illustrated in figure 1.

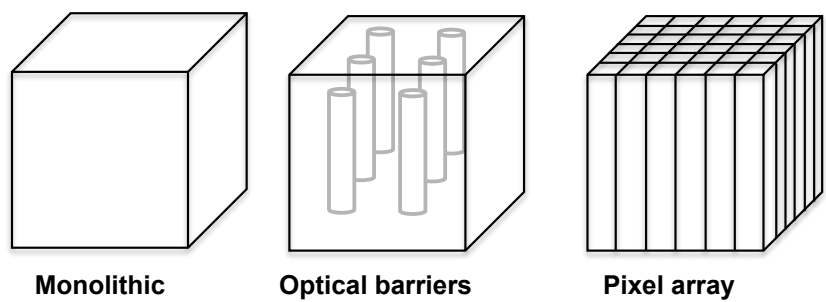

**Figure 1**. Schematic figure showing how a detector fabricated using the LIOB technique is positioned between the two extremes of a monolithic detector block and a mechanically pixelated array. The optical barrier pattern shown in this figure is just an example aimed to illustrate the flexibility in the shapes that can be introduced into the crystal using the LIOB technique.

Among the huge barrier pattern space available thanks to the flexibility of the LIOB technique, in this work we only study the light transport in detectors with simple pixel-like optical barrier patterns, extending all the way, or half way through a LYSO:Ce crystal. We then compare the expected detector performance of these laser-processed detectors with that of a monolithic crystal and a pixelated array. Given that the optical barriers allow for some amount of cross talk between pixels, the light spread function of a detector with optical barriers has the potential to combine the DOI dependency inherent to a monolithic detector with the transversal resolution of a pixelated array.

## 2. Materials and methods

For all simulations presented here we have used the Monte Carlo code DETECT2000 (Cayouette *et al* 2003) to simulate the scintillation light transport in the detectors. In the following sections, we describe the different detector types and configurations that were implemented, the simulation parameters used to model the detectors, and finally how the simulations were set up in order to study and compare the performance of each of the detector types.

*2.1. Detector types*

We have studied three scintillator detector types: monolithic, mechanically pixelated arrays and laser-processed scintillators containing optical barriers. As discussed in the introduction of this paper, one can envisage the first two types as two distinct categories, and the latter as a new category combining features of the other two.

In all cases the total scintillator detector dimension was kept constant at 25.4x25.4x20.0 mm³ in order to match the cross-section of the photodetector array described in section 2.3. Furthermore, in all cases the side- and entrance surfaces of the detector were wrapped in an external diffuse reflector with a reflection coefficient of 0.98, corresponding to 3 layers of Teflon tape (Janecek and Moses 2008).

The mechanical array was simulated as 21x21 individual crystals, each with a dimension of 1.0x1.0x20.0 mm³, individually wrapped in Teflon tape. The dead space between pixels was set to 0.2 mm, and no photon transport was modeled in the reflector material. The roughness of the outer surfaces of the monolithic crystal as well as each individual pixel in the mechanically pixelated array was varied.

For the laser-processed detectors the outer crystal surface was kept polished, and the roughness of the barrier-crystal interface as well as the barrier RI were varied. The two studied laser-processed detector configurations are shown in figure 2. They correspond to a detector with optical barriers in a pixel-like pattern extending all the way through the crystal thickness (a) and barriers only in the top half of the crystal (b). In both cases the barriers are arranged with 1 mm separation forming a 24x24 array where each pixel-like volume has a cross-section of 1.05x1.05 mm².

The mechanically pixelated array and the laser-processed geometry (a) were simulated with and without a 1 mm thick light guide (RI=1.5), while the rest of the configurations were simulated without light guide. In detector configurations that rely on optical isolation between scintillator pixels a light guide is needed to spread the scintillation light over multiple photodetector pixels for accurate event positioning, unless one-to-one coupling between crystal pixel and photodetector pixel is employed. This spread is inherent in a monolithic detector, where a light guide would only serve as a mean to remove the possibility of gamma ray interactions very close to the photodetector plane.

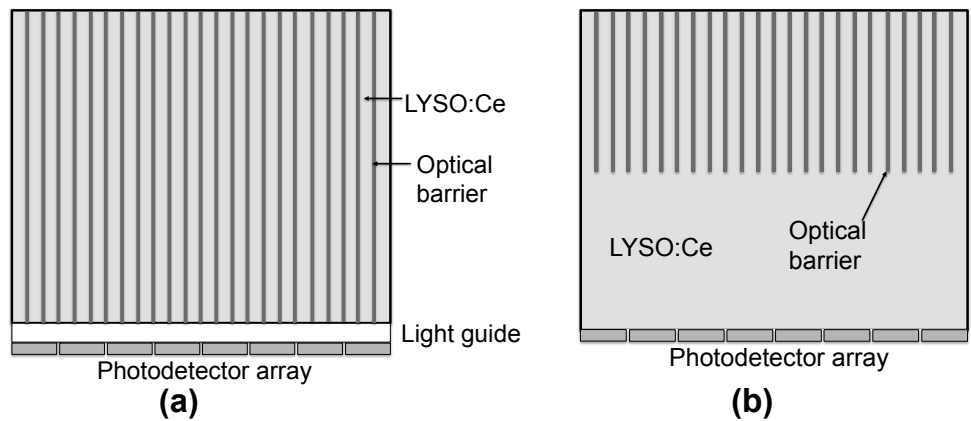

**Figure 2.** Schematic side views of the laser-processed detector configurations. In (a) the optical barrier pixel pattern extends all the way through the 20 mm crystal thickness, and in (b) only the top half (10 mm) is processed. Geometry (a) is shown with light guide, but was also simulated without light guide.

*2.2. LYSO:Ce crystal*
The LYSO:Ce crystal was simulated with a refractive index of 1.82 at 420 nm. An optical absorption length of 40 cm was used, and scattering of the optical photons within the crystal bulk was not considered (i.e. infinite scatter length). These values were chosen based on a literature survey where no clear consensus was found on the best values to choose, and the properties have also been shown to vary between crystals from different manufacturers (Steinbach *et al* 2012). Furthermore, the attenuation is often measured as a single quantity without separating the components of scattering and absorption. Given that our detector modules are relatively small, scattering should not have a significant impact and the absorption length used is found within the range of values used and measured by others (Berg *et al* 2015, García *et al* 2007, Ogata *et al* 2014, van der Laan *et al* 2010, Zhang *et al* 2013).

*2.3. Photodetector array*
We used a simplified model of Hamamatsu S13361-3050AE-08 Multi-Pixel Photon Counter (MPPC). This device is an 8x8 MPPC array of 3.0x3.0mm² pixels with 3.2 mm pixel pitch. The array has a 0.1 mm thick entrance window with a refractive index of 1.55 (Hamamatsu Photonics K.K. 2016). In our simulations, the optical photons intersecting with the active area of the MPPC pixels were treated as "counted" and those reaching the inter-pixel dead space as "lost". This setup does not take into account reflections from the dead space or the active surface back into the entrance window and scintillator crystal, but captures the transversal distribution of photons impinging on the photodetector plane.

*2.4. Optical barriers*
The optical barriers were modeled as 50 μm thick slabs in all simulations reported in this paper. This barrier thickness is based on experimental observations of barrier thicknesses between approximately 20 to 50 μm in LYSO:Ce that was previously reported (Sabet *et al* 2012a). The size depends on the laser parameters and delivery optics used during the process, as well as on the crystal material. It is noteworthy that no significant difference in light confinement was observed when varying the barrier thickness within this range in the simulations. It should be noted that DETECT2000 is based only on a geometrical optics model (aka ray optics). If the barrier size becomes comparable to the emission wavelength of LYSO:Ce, a wave optics model should be used to study the light transport in the crystal since diffraction and interference can become significant components of the light transport.

The refractive index of the barriers also depends on the laser parameters used during processing, where the lowest achievable RI is 1.0, corresponding to void formation in the crystal. In this work, we have simulated a range of barrier RI values between 1.0 and 1.6.

The interface between the optical barrier and the unmodified crystal bulk was described using the POLISH and the UNIFIED surface models implemented in DETECT2000. POLISH corresponds to a perfectly smooth interface where all reflections and refractions are specular around the nominal surface normal. The UNIFIED surface model can be used to describe a range of surface roughness values between a specular and completely diffuse interface, and also linear combinations between different types of reflections (Levin and Moisan 1996). In this work we have used specular lobe reflection characterized by the $\sigma_\alpha$ parameter, which is recommended for simulations of a rough interface between two dielectric surfaces (Moisan *et al* 2000). The $\sigma_\alpha$ parameter corresponds to the standard deviation of a Gaussian distribution of surface normals around the nominal surface normal, and is typically determined by constraining it to surface roughness data. For an interface like the one between an optical barrier and the crystal bulk there is no straightforward way of obtaining such data since the barriers are contained inside the crystal bulk, and placing them close to an edge for easier characterization will affect their properties. Furthermore, in earlier work, we showed that the barrier roughness can be controlled to some extent by varying laser parameters (Bläckberg *et al* 2016a). For this work we therefore simulated a range of surface roughness values, and plan to compare the results to experimental measurements in the future.

*2.5. Simulation setup*

For each of the detector types described in section 2.1 we generated one dimensional light response functions (LRFs), position histograms and line profiles as a function of gamma-ray DOI. We also extracted the light collection efficiency for all detector types. Only the light spread from photopeak events where the full gamma-ray energy is absorbed in the first interaction was simulated. All simulations were started with different initial seeds, making them statistically independent.

*2.5.1. Light response function.* The procedure to generate the LRF as a function of DOI consists of beam scans along one central pixel row, as illustrated in figure 3. An isotropic source of 420 nm optical photons was placed in the center of each crystal pixel (or with 1 mm separation for the monolithic crystal) and each photon was tracked until termination. The procedure was repeated at different crystal depths with 3 mm separation, and the number of photons detected by each MPPC pixel was recorded as a function of source location. The results were used to generate 1D LRFs as a function of DOI for each of the detector configurations, as well as for determination of the light collection efficiency, as further described in section 2.5.3.

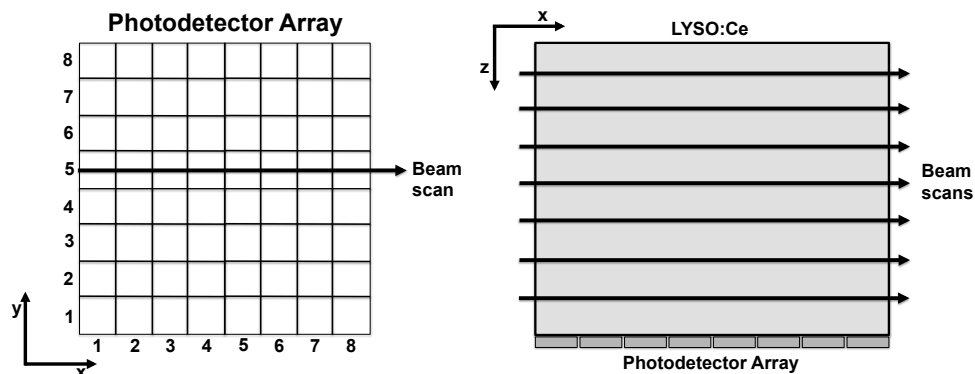

**Figure 3.** Illustrations of the beam scans performed for generation of light response functions. Left: XY-view, right: XZ view.

*2.5.2. Position histogram and line profile.* To generate position histograms and corresponding line profiles for each configuration, we performed beam scans with 200 gamma events simulated at each location. LYSO:Ce has a light yield of 27-32 photons per keV, which yields about 13800-16300 photons for photoelectric absorption of 511 keV gammas. The S13361-3050AE MPPC series have ~37% photon detection efficiency (PDE) at 420 nm. Therefore we started the light transport simulation with 5000 photons per gamma-ray interaction and set the quantum efficiency for the modeled photodetector to 1. We used a simple centroid event positioning estimator on the MPPC signals to generate position histograms and their associated line profiles. While this simple angler logic is not typically used for event positioning in monolithic crystals, but we chose this method to make a simple and straightforward comparison between the different detector configurations. Nowadays more sophisticated algorithms, such as Maximum Likelihood (ML) are being used, which could be very well suited for the laser-processed detectors and this is something we aim to explore in future work.

*2.5.3. Light collection efficiency.* The light collection efficiency for each detector configuration was determined by averaging the total number of counted photons over all interaction locations simulated for the light response functions. Information regarding where losses occurred (i.e. bulk absorption, trapping, or surface losses) could also be extracted from this data.

## 3. Results

While we simulated a range of barrier RI values, the majority of the presented results will be for RI=1.0 corresponding to that of the void as the best-case scenario, and for RI=1.4 as a mid-range value between void and the crystal bulk. In the following sections, LRFs, position histograms and line profiles are presented for each of the three scintillator categories, including the two variations of the laser-processed detector. The light collection efficiencies for all detector configurations are summarized in section 3.4 In order to make the figures more informative, LRFs for events very close to the photodetector plane (i.e. z=19 mm), with very large solid angle, are omitted. In all cases the LRFs were generated by normalization to the number of emitted photons from each source location.

### 3.1. Monolithic detector

The monolithic detector was simulated with varying surface roughness. Figure 4 shows the resulting LRF for a polished crystal, and for one with a rough outer surface characterized by $\sigma_\alpha=20°$. In both cases the LRF changes as a function of DOI. Furthermore the LRF is nearly flat for events taking place in the top part of the crystal. Figure 5 shows the FWHM width of the LRF as a function of DOI and crystal surface roughness. The values were obtained by fitting with a Gaussian function. In the half of the crystal closest to the photodetector plane the depth dependence is almost linear with decreasing width for increasing DOI, especially for the polished crystal. In the crystal half further away from the photodetector plane, fitting becomes difficult due to the flat LRFs that are heavily affected by reflections from the detector sides. Figure 6 shows position histograms and line profiles for a polished crystal, as well as a rough one with $\sigma_\alpha=20°$. Also in this figure the depth dependence of the detector response is apparent, with better position separation close to the photodetector. It is also evident that the flat nature of the LRF, especially close to the detector entrance surface, makes event positioning using the simple centroid estimator difficult. Furthermore, a rough outer detector surface results in narrower line profiles and impaired transversal detector resolution compared to a polished crystal.

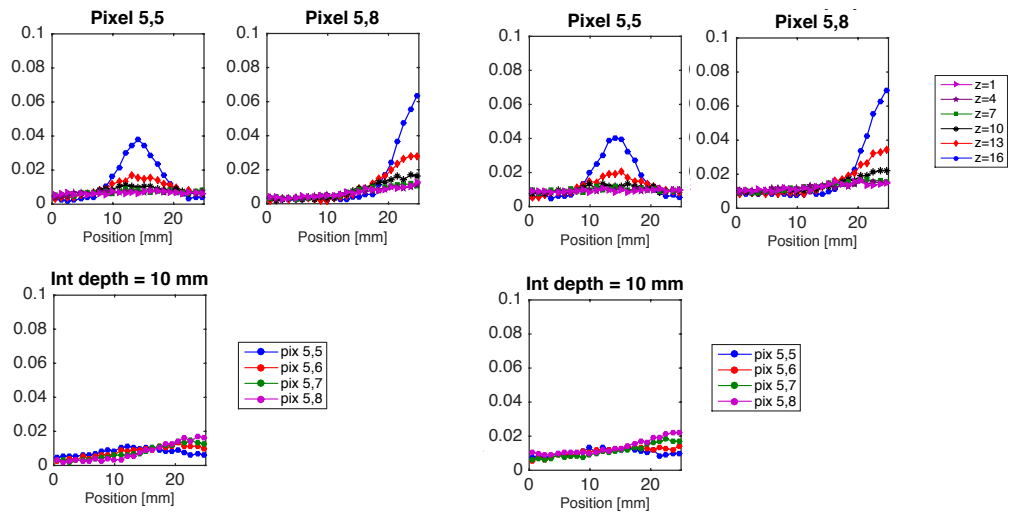

**Figure 4.** LRF for the monolithic detector as a function of optical photon source location. Z increases with DOI. Left: The outer crystal surfaces are polished. Right: The outer crystal surfaces are rough with $\sigma_\alpha=20°$. For each configuration the LRF of one central and one edge MPPC pixel are shown as a function of DOI (top row), as well as the LRF at 10 mm interaction depth for four adjacent MPPC pixels (bottom row).

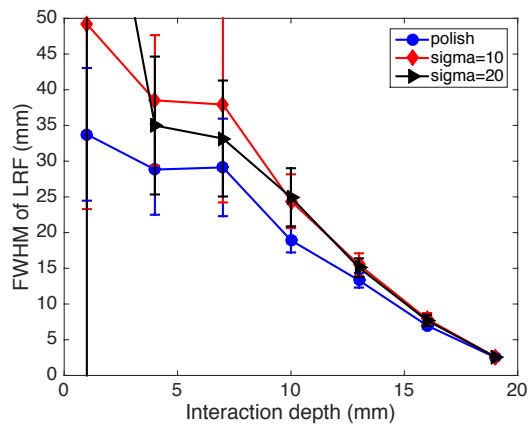

**Figure 5.** FWHM of the LRF for one central MPPC pixel as a function of DOI and surface roughness. The values are obtained by fitting the LRF with a Gaussian curve and determination of the FWHM width. The error bars correspond to the 95% confidence interval of the fitting parameter.

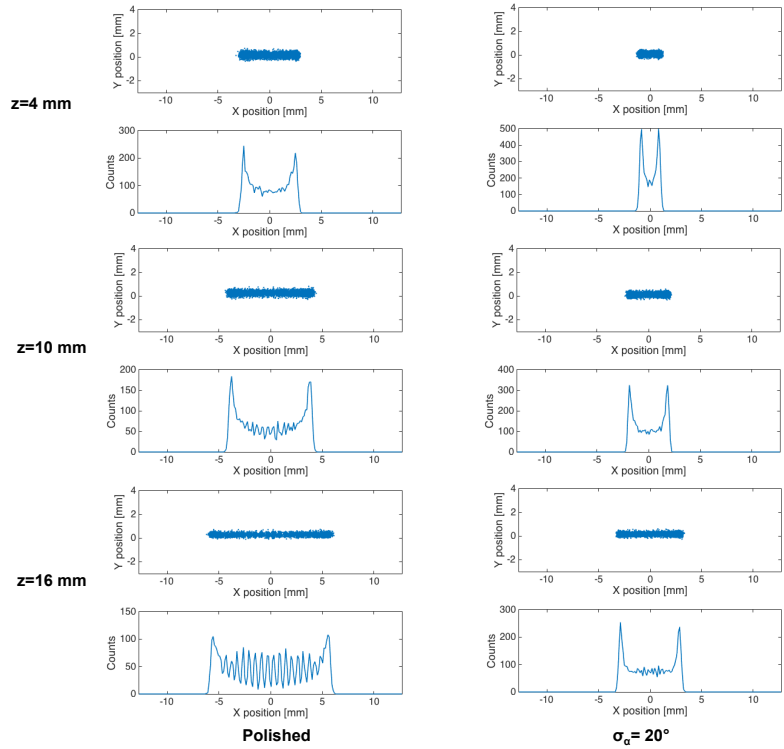

**Figure 6.** Position histograms and line profiles for the monolithic detector as a function of surface roughness and DOI.

*3.2. Mechanically pixelated array*

Figure 7 shows the corresponding LRFs for a mechanically pixelated array, with and without light guide. For this detector type there is no DOI dependence observed in the LRF when the scintillator pixel surfaces are polished. With a rough pixel surface, one only sees a very marginal dependence that may not be useful to extract any DOI information. The FWHM curves in figure 8 show that the width of the LRF is independent of DOI, regardless of surface roughness of the pixels, and whether or not a light guide is used. As expected, the light guide makes the LRF slightly wider. The line profiles in figure 9 show that for a pixel array without a light guide, it is not possible to separate events taking place in crystal pixels located above the same MPPC pixel since their scintillation light is collected by only one MPPC and therefore other MPPC signals cannot contribute to accurate event positioning. As expected, by using a light guide, the scintillation light will spread over multiple MPPC pixels and the crystal pixels can then be separated. In figure 9, one can see that a rough pixel surface helps in pulling the pixels away from each other in the position histogram. It does however not have any effect on DOI information.

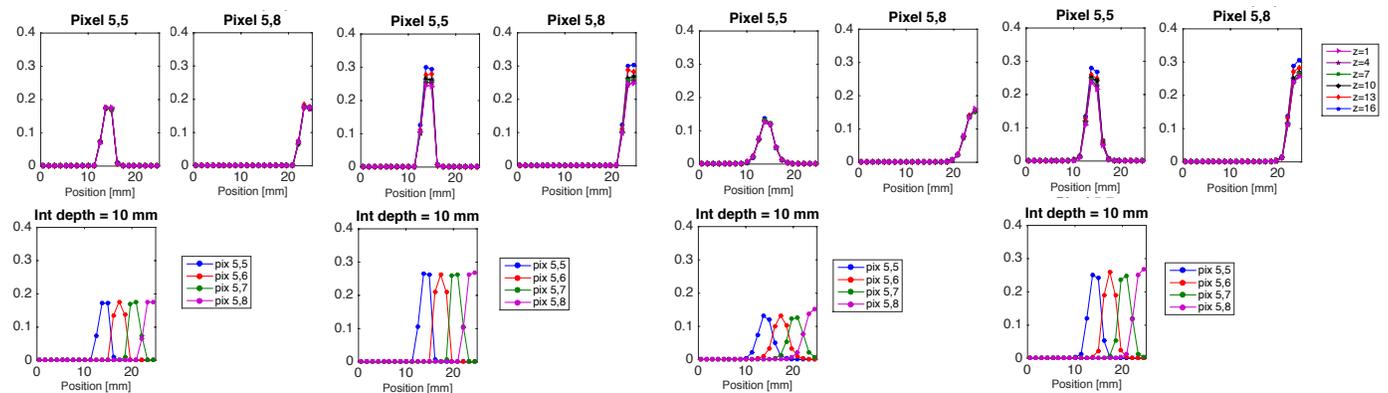

**Figure 7.** LRF for a mechanically pixelated array as a function of DOI and pixel surface roughness. From left to right: Without light guide and polished sides, without light guide and rough sides with $\sigma_\alpha=20°$, with light guide and polished sides, with light guide and rough sides with $\sigma_\alpha=20°$. For each configuration the LRF of one central and one edge MPPC pixel is shown as a function of DOI (top row), as well as the LRF at 10 mm interaction depth for four adjacent MPPC pixels (bottom row).

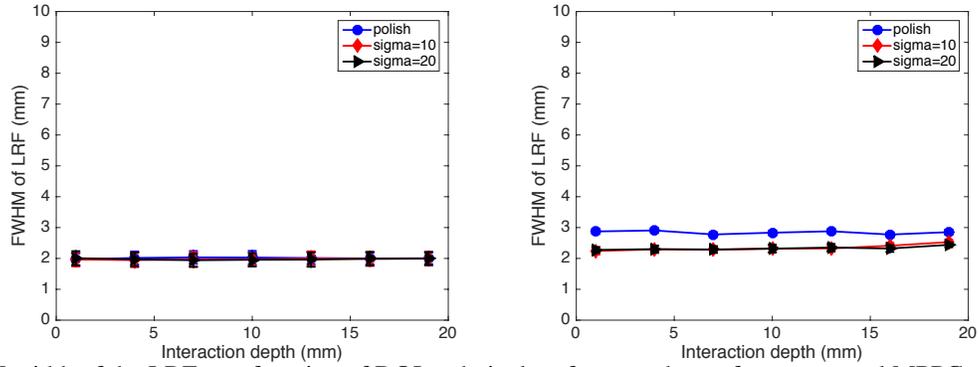

**Figure 8**. FWHM width of the LRF as a function of DOI and pixel surface roughness for one central MPPC pixel. The values are obtained by fitting the LRF with a Gaussian curve. The error bars correspond to the 95% confidence interval of the fitting parameter. Left: Without light guide, Right: With a 1 mm light guide.

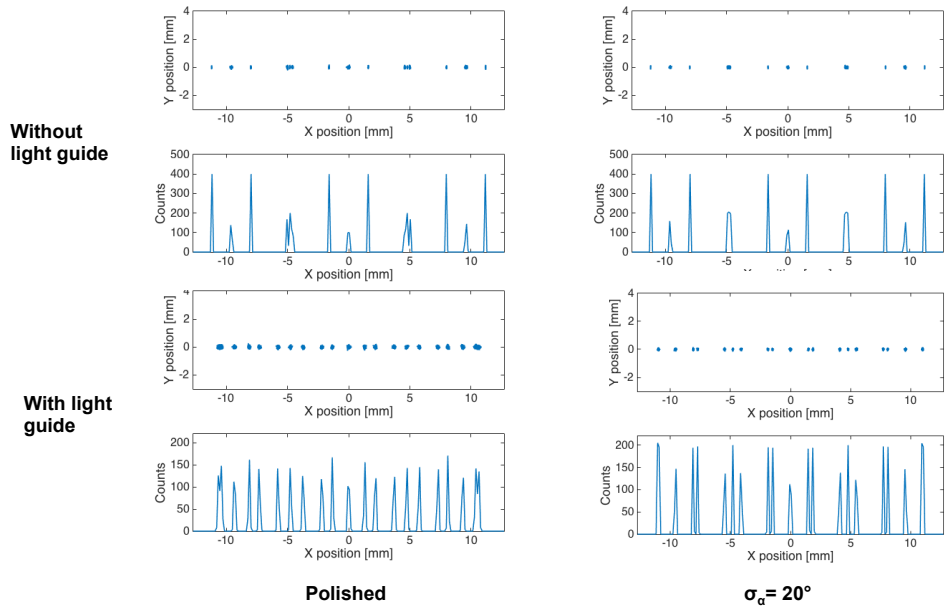

**Figure 9.** Position histograms and line profiles for a mechanically pixelated array. All plots show events at central depth only, since as seen in figure 7 there is no significant depth dependence of the LRF for this detector type.

*3.3. Laser-processed detectors*

3.3.1. *Optical barriers all the way through the crystal thickness.* Figure 10 shows the resulting LRFs for smooth barrier-crystal interfaces and barrier RI=1.0, with and without light guide. These results are similar to the mechanically pixelated array and no depth dependence of the LRF can be observed in either case. While the use of a light guide helps spreading the scintillation light over multiple MPPC pixels, they do not contribute to DOI, therefore in the remainder of the paper, we will present results without light guide for the purpose of simplicity and more straightforward comparison to the partially processed detector.

Figure 11 and figure 12 demonstrate how the roughness of the barrier-crystal interface affects the LRF for barrier RI=1.0 and RI=1.4, respectively. In both cases increased interface roughness enhances the depth dependence of the LRF. This dependence is further analyzed in figure 13 where the FWHM width of the LRF for one central MPPC pixel is shown as a function of DOI and interface roughness. It can be seen that for a smooth barrier-crystal interface the width of the LRF is independent of interaction depth, while with increasing interface roughness, the LRF becomes narrower with increasing DOI.

Figure 14 shows the FWHM of the LRF as a function of DOI and barrier refractive index for barriers with a smooth barrier-crystal interface as well as a rough interface with $\sigma_a$=20°. As previously seen, for a smooth interface there is no depth dependence, while for the rougher interface the width of the LRF depends on the interaction depth. This dependence becomes more pronounced when the barrier RI approaches that of the LYSO:Ce crystal, in which case the behavior of the detector will approach that of a monolithic one.

Finally, figure 15 shows position histograms and line profiles produced using the centroid estimator for different RI-interface combinations. For the barrier-crystal interface characterized by $\sigma_\alpha=20°$ all 24 pixels may be resolved, while for the smoother interface crystal pixels positioned above the same MPPC pixel are merged. This trend is seen for both barrier RI=1.0 and barrier RI=1.4.

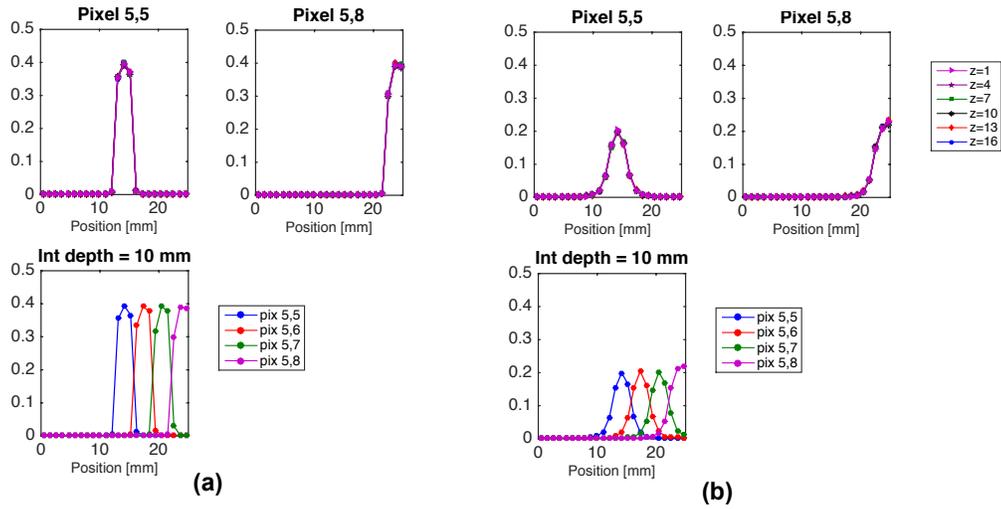

**Figure 10.** LRFs for barriers all the way through the crystal thickness. The results are here shown for barrier RI=1.0 and a perfectly smooth barrier-crystal interface. (a) without a light guide, (b) with a 1 mm thick light guide. For each configuration the LRFs of one central and one edge MPPC pixel are shown as a function of DOI (top row), as well as LRFs at 10 mm interaction depth for four adjacent MPPC pixels (bottom row).

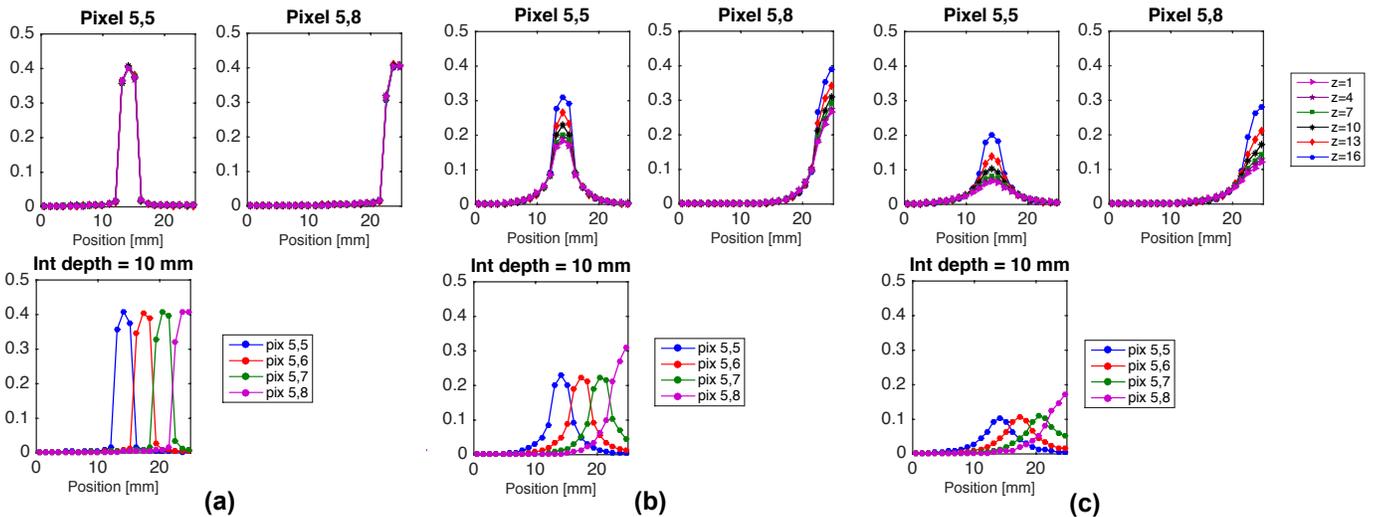

**Figure 11.** LRF as a function of depth for varying surface roughness. In all cases the barrier RI=1.0 and the barrier-crystal interface is varied using the UNIFIED surface model with varying $\sigma_\alpha$ parameter. (a) $\sigma_\alpha=1°$, (b) $\sigma_\alpha=20°$, (c) $\sigma_\alpha=60°$. For each configuration the LRFs of one central and one edge MPPC pixel are shown as a function of DOI (top row), as well as LRFs at 10 mm interaction depth for four adjacent MPPC pixels (bottom row).

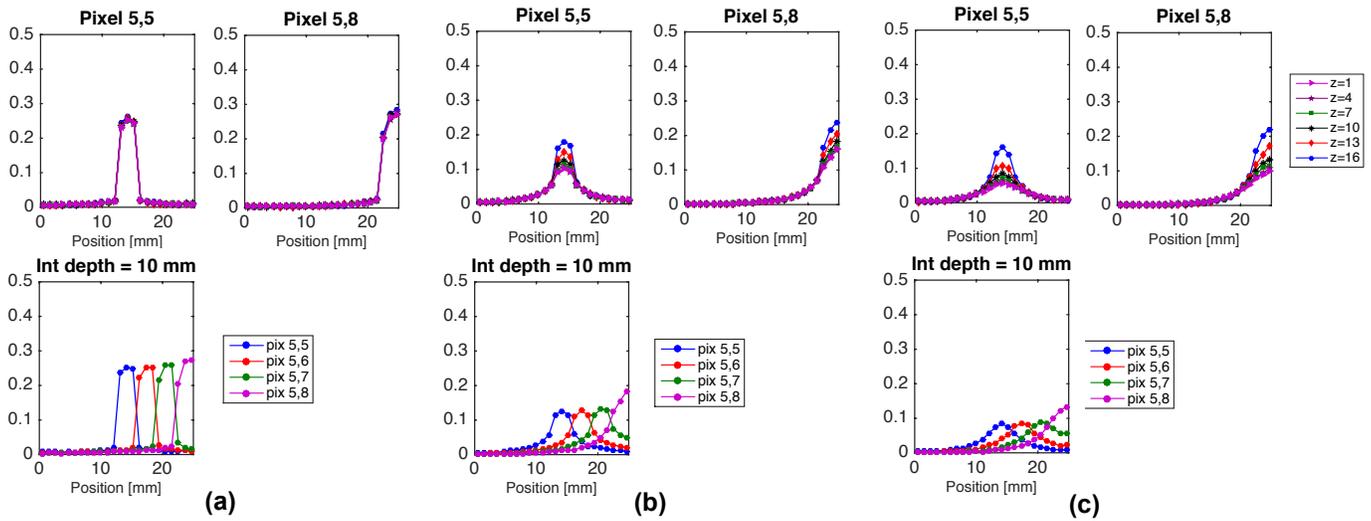

**Figure 12.** Same as figure 11 but with barrier RI=1.4.

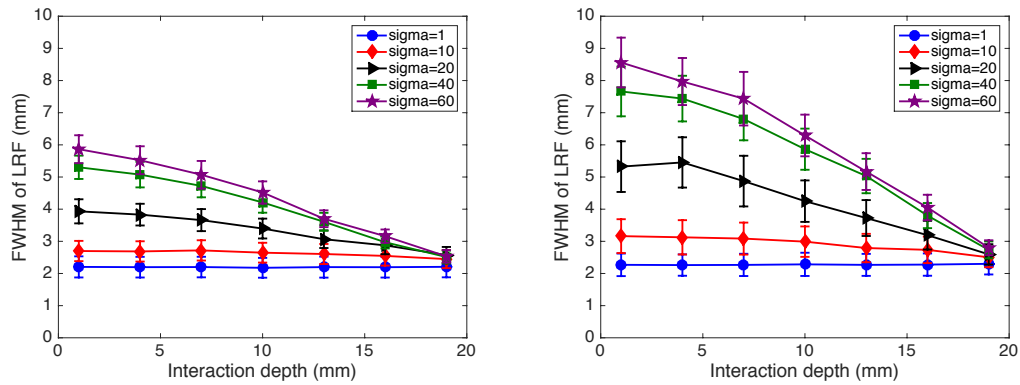

**Figure 13.** FWHM of the LRF for one central MPPC pixel obtained by fitting with a Gaussian curve. The error bars correspond to the 95% confidence interval of the fitting parameter. FWHM values are shown as a function of interaction depth and barrier-crystal interface roughness. The roughness is varied through the $\sigma_\alpha$ parameter in the UNIFIED surface model. Left: Barrier RI=1.0, right: barrier RI=1.4.

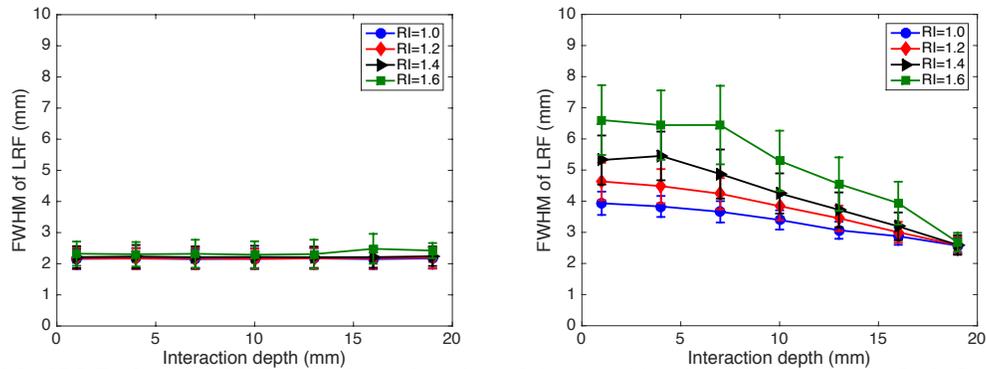

**Figure 14.** FWHM of LRF of central MPPC pixel as a function of depth and barrier refractive index. Left: Polished barrier-crystal interface. Right: Roughened barrier-crystal interface characterized by $\sigma_\alpha$=20°.

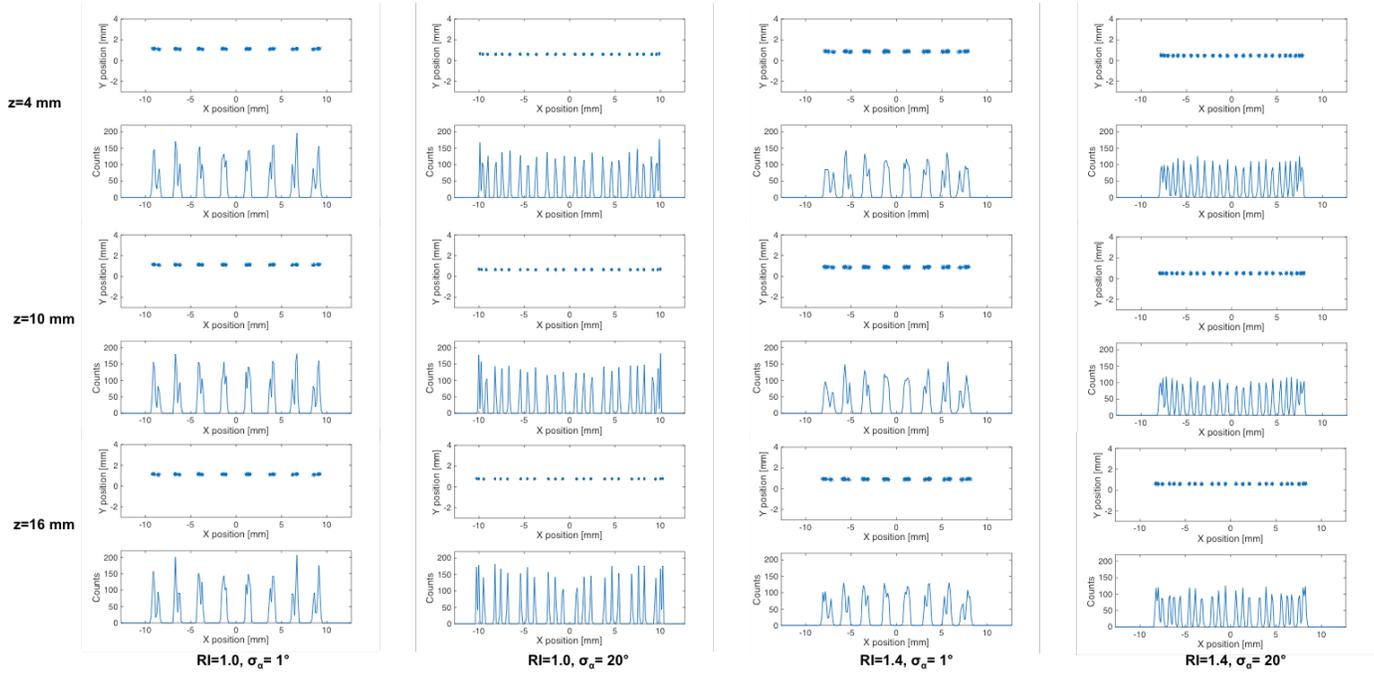

**Figure 15.** Position histograms and corresponding line profiles for a laser-processed detector with optical barriers extending all the way through the crystal thickness. From left to right: RI=1.0 and barrier crystal interface with $\sigma_\alpha$=1°, RI=1.0 and barrier crystal interface with $\sigma_\alpha$=20°, RI=1.4 and barrier crystal interface with $\sigma_\alpha$=1°, RI=1.4 and barrier crystal interface with $\sigma_\alpha$=20°.

*3.3.2. Barriers in top 10 mm.* Figure 16 and figure 17 shows the LRF for a detector with optical barriers half way through the crystal thickness, as a function of DOI and barrier-crystal interface roughness, for barrier RI=1.0 and RI=1.4, respectively. It can be observed that the LRF from events within the pixelated region become narrower as the interface roughness is increased, indicating that some DOI information could be extracted from the LRF, especially in the case of a rougher interface. This behavior is also seen in the FWHM curves shown in figure 18 and figure 19. Compared to the case with optical barriers all the way through the crystal thickness these curves have a more complex behavior caused by having different detector configurations in the top and bottom half of the scintillator crystal. Also for this barrier pattern a strong dependence of the LRF on the roughness of the barrier-crystal interface can be seen. Finally, figure 20 shows position histograms and line profiles, as a function of DOI, interface roughness, and barrier RI. In all cases the central pixels are well resolved while the side pixels are merged, and the performance is in general better with a rougher interface. There is also an apparent depth dependent behavior in the line profiles, where the pixel separation is larger for gamma-ray interactions closer to the entrance surface of the detector. The same trends can be seen regardless of barrier RI, but the line profiles are slightly squeezed for the higher barrier RI.

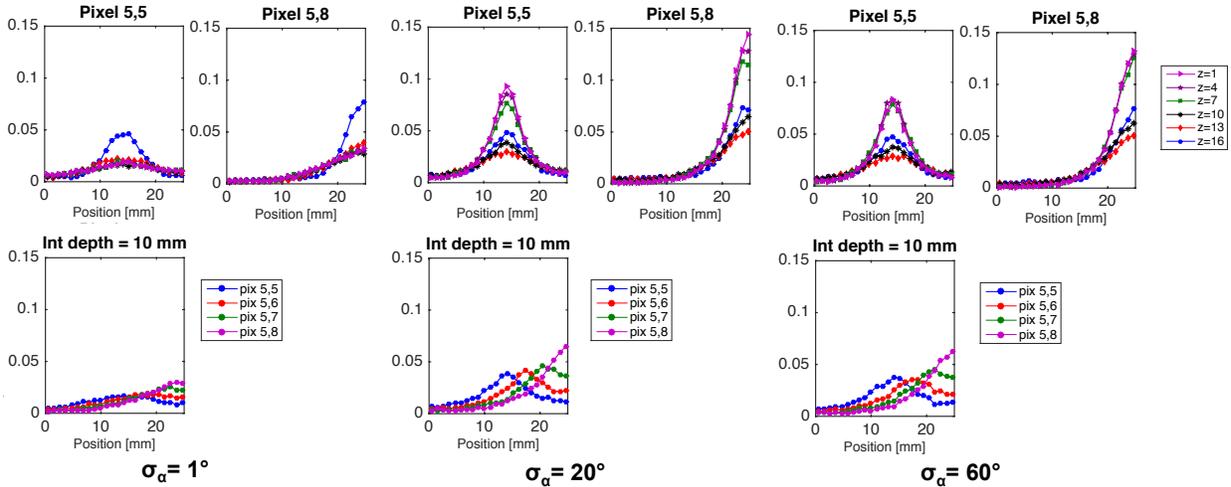

**Figure 16.** LRF for a laser-processed detector with optical barriers in the top half of the crystal. The results are shown as a function of DOI for varying roughness of the barrier-crystal interface. The barrier RI=1.0 in all three cases. For each configuration the LRF of one central and one edge MPPC pixel are shown as a function of DOI (top row), as well as the LRF at 10 mm interaction depth for four adjacent MPPC pixels (bottom row).

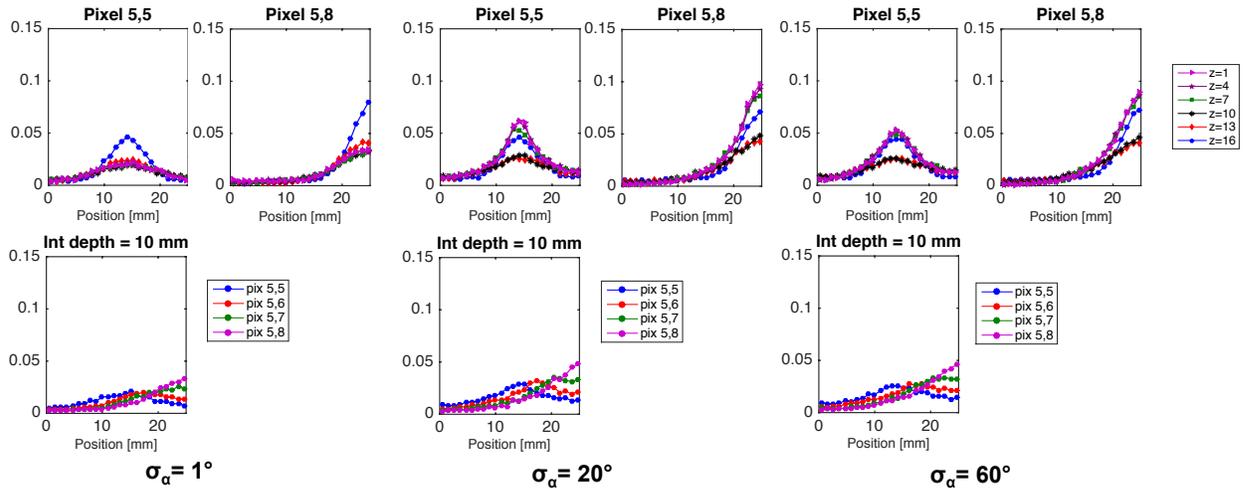

**Figure 17.** Same as figure 16, but with barrier RI=1.4.

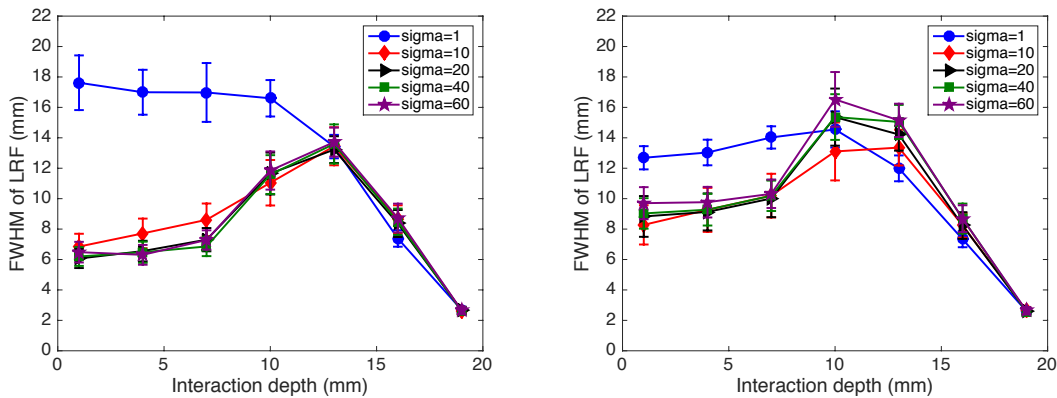

**Figure 18.** FWHM values obtained by fitting the LRF with a Gaussian function, shown as a function of DOI and barrier-crystal interface roughness. The error bars correspond to the 95% confidence interval of the fitting parameter. Left: Barrier RI=1.0, Right: Barrier RI=1.4.

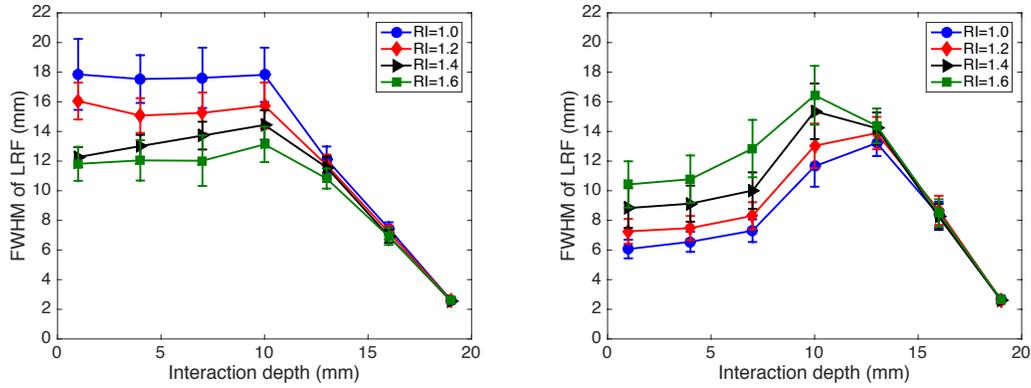

**Figure 19.** FWHM values obtained by fitting the LRF with a Gaussian function, shown as a function of DOI and barrier RI. The error bars correspond to the 95% confidence interval of the fitting parameter. Left: Barrier-crystal interface is polished, Right: Barrier-crystal interface is rough with $\sigma_\alpha=20°$.

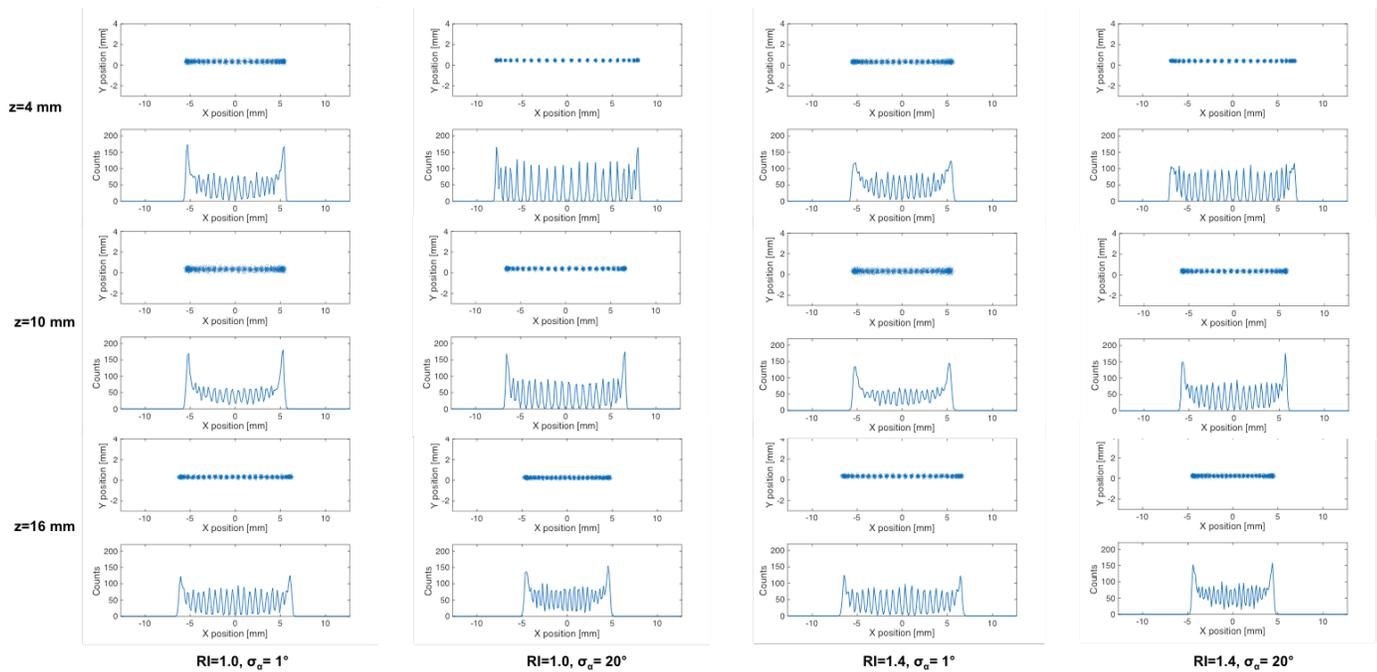

**Figure 20.** Position histograms and corresponding line profiles for a laser-processed detector with optical barriers extending half way through the crystal thickness. From left to right: RI=1.0 and barrier crystal interface with $\sigma_\alpha=1°$, RI=1.0 and barrier crystal interface with $\sigma_\alpha=20°$, RI=1.4 and barrier crystal interface with $\sigma_\alpha=1°$, RI=1.4 and barrier crystal interface with $\sigma_\alpha=20°$.

*3.4. Light collection efficiency*
Table 1 summarizes the light collection efficiency and the nature of the photon losses in each of the simulated detector configurations. The numbers quoted are the average over all start positions in the beam scans described in section 2.5. One can see that the light collection efficiency is increased for detectors containing optical barriers compared to both the monolithic detector block and the mechanically pixelated array. Furthermore, it can be noted that the losses for a monolithic detector, as well as a laser-processed one, are dominated by optical absorption in the crystal bulk, while for the mechanical array the losses occur mainly at surfaces due to non-perfect reflectors. The reason is a dramatic increase in surface reflections in the high aspect ratio pixels compared to the monolithic detector block. Shown in the table is also how the light collection efficiency varies over the detector volume, and noteworthy is that the uniformity of the light collection efficiency is better in the monolithic detector compared to the mechanical array. For the laser-processed detectors the configuration with optical barriers all way through the crystal thickness has more uniform light collection efficiency compared to the configuration with barriers only half way through the crystal.

**Table 1.** Light collection efficiency and types of losses in the studied detector configurations. "Other losses" refer to photons being absorbed at reflectors, hitting the dead space of the photodetector array, escaping the geometry or being internally trapped. The two latter are minor in all cases. For the monolithic detector and the pixelated array "Polish" and $\sigma_\alpha = 20°$ corresponds to the outer surfaces of the crystal block as well as the individual pixels. For the geometries containing optical barriers the outer crystal

surface is polished in all cases and surface roughness specifications correspond to the barrier-crystal interface. All results shown are without light guide. The uncertainties given in the "Counted" columns reflect the standard deviation in the number of counted photons over all gamma-ray interaction points used to calculate the average light collection efficiency.

|                    | Polished     |              |           | $\sigma_\alpha = 20°$ |              |           |
| ------------------ | ------------ | ------------ | --------- | --------------------- | ------------ | --------- |
|                    | Counted (%)  | Absorbed (%) | Other (%) | Counted (%)           | Absorbed (%) | Other (%) |
| Monolithic         | 39.4 ± 0.5   | 48.5         | 12.1      | 71.5 ± 0.9            | 17.8         | 10.7      |
| Mechanical array   | 33.0 ± 2.9   | 27.6         | 39.4      | 51.2 ± 7.2            | 12.7         | 36.1      |
| OB all way, RI=1.0 | 44.3 ± 2.3   | 42.4         | 13.3      | 74.2 ± 2.2            | 13.6         | 12.2      |
| OB all way, RI=1.4 | 45.9 ± 2.1   | 39.9         | 14.2      | 76.6 ± 1.0            | 10.9         | 12.5      |
| OB 10 mm, RI=1.0   | 44.9 ± 1.0   | 43.4         | 11.7      | 72.9 ± 3.7            | 15.5         | 11.6      |
| OB 10 mm, RI=1.4   | 46.1 ± 1.1   | 41.5         | 12.4      | 73.0 ± 4.3            | 15.2         | 11.8      |

## 4. Discussion and conclusions

Through our simulations, we have shown that the behavior of a laser-processed LYSO:Ce detector containing optical barriers falls between that of the two extreme detector types, monolithic and mechanically pixelated arrays. The behavior of a laser-processed crystal can be tailored to be closer to a monolithic crystal or a pixelated detector array, thanks to the huge laser parameter space as well as barrier pattern space that the LIOB technique can offer. Among countless number of optical barrier patterns, we selected two simplified versions that are easy to implement experimentally and easy to follow conceptually. The all-the-way barrier pattern resembles the mechanically pixelated array type and the top-half barrier pattern is similar to work that has been presented elsewhere (Kaul et al 2013, Gonzalez-Montoro et al 2017).

We first simulated the monolithic and the pixelated array detector types in DETECT2000, as these constitute the mainstay in imaging applications and have been in use for decades. Even though other structured scintillator types have been developed, such as microcolumnar detectors (Nagarkar et al 1998, Sabet et al 2012b), and scintillators that are grown into pixel shapes (Sabet et al 2013, Zhao et al 2015), we focused this study on the most common detector types for nuclear medicine applications. These initial simulations helped us debug our code and setup baselines for the new detector designs fabricated using the LIOB technique. The results presented in figures 4-6 represents the LRF, line profile and position histogram of a monolithic LYSO:Ce detector. As expected, the LRF and its width is a function of gamma-ray interaction depth, and our results are in good agreement with those reported elsewhere (Kaul et al 2013, Ling et al 2007a, Lerche et al 2005, Tavernier et al 2005). We also notice that the position histogram and line profile is more expanded when using a polished monolithic crystal compared with a crystal with a rougher surface finish. The LRF width has a linear behavior at interaction depths greater than 13 mm, which bodes well with the fact that thin monolithic crystals are great choice when DOI is needed and very high transverse spatial resolution is not demanded. As expected, the detector spatial resolution is degraded near the edge area due to the so-called edge effect that is more pronounced in these detector types when using a centroid positioning estimator, compared to pixelated detectors.

The corresponding results for a pixelated detector array are shown in figures 7-9. Here no DOI is achievable with single-side readout and the centroid positioning algorithm. The detector transverse resolution is better than in a monolithic detector, and all pixels in the array are resolved in the position histograms and line profiles when a light guide is used. Note that the sharp slope of the LRFs in the bottom row of figure 7 manifests this superior transverse resolution when using the centroid positioning algorithm. However, it is also apparent that without light guide the scintillator pixels cannot be resolved (see figure 9). The high transversal resolution makes this detector type the backbone of the majority of the high spatial resolution imagers where the detector thickness can be increased to enhance the system sensitivity with little effect on the resolution. On the other hand, given that no DOI can be extracted, a thick detector may lead to image blurring in the FOV periphery, especially in small animal PET systems. The laser-processed detector with optical barriers all the way through the crystal thickness visually resembles the pixelated array type, but shows a behavior that is a combination of that of the monolithic detector and the pixelated array, depending on the barrier properties. In figure 10, we observe that when the barrier RI equals that of air (RI=1.0) and the barrier-crystal interface is smooth, no DOI information can be extracted, regardless of existence of a light guide. A light guide will, however, spread the light over multiple MPPC pixels and can be used to avoid merging pixels in the flood maps. This behavior can be most useful when high light channeling in each of the pixel-like elements is demanded, for example in photon counting CT detectors. Figures 11 and 12 show the LRF plots for the same detector type with rough barrier-crystal interface and barrier RI of 1.0 and 1.4, respectively. Note that in both figures, we see a depth dependency of the LRF as the barrier-crystal interface becomes rougher. The DOI behavior is clearly demonstrated in figure 13 showing that a continuously linear DOI can be extracted from the width of the LRF when the barrier-crystal interface is rough. However, the results presented in figure 14 suggest that when the barrier-crystal interface is smooth, the width of the LRF is independent of DOI, regardless of the value of the barrier RI. Position histograms and their associated line profiles at 3 crystal depths (figure 15) show that excellent transverse spatial resolution can be achieved with a rough barrier-crystal interface. The transverse resolution is superior when the barrier RI equals 1.0.

In the detector with optical barriers only in the top half of the crystal, the LRF gives subtle DOI information when the barrier-crystal interface is smooth. However the DOI dependency of the LRF becomes stronger when this interface is rough, following a similar trend compared with the detector with all-the-way barrier pattern (see figures 16 & 17). In figure 18, we observe a complex DOI response in that when the barrier RI is 1.0 and the barrier-crystal interface is rough, the LRF width linearly increases with DOI until a tipping point at 13 mm after which it starts to decrease sharply with interaction depth. With RI of 1.4, there is larger light spread as expected but the tipping point in the plot with different barrier-crystal interface roughness takes place at a crystal depth of 10 mm. In the two linear areas of the LRF width curve (before and after 13 mm crystal depth for barrier RI=1.0), we can observe that for example the LRF width is the same for interaction depths of 7 and 13 mm. However, as can be seen in figure 16, the LRF signals of the MPPCs have different values for these two cases, which can be used to distinguish between gamma-rays interacting at 7 or 13 mm depths. Figure 19 shows that when the barrier-crystal interface is rough, a wide range of barrier RIs can be used to extract multiple DOI levels, a flexibility that can be used to fine-tune the transverse vs DOI resolution. Note that with smooth barrier-crystal interface, the LRF width demonstrates a similar trend compared to a pixelated detector or an all-the-way barrier detector for events in the part of the crystal containing optical barriers, but a behavior similar to a monolithic detector in the unprocessed part of the crystal. This is true regardless of the barrier RI. Figure 20 shows that compared to a monolithic detector, the transverse resolution can be enhanced while degrading the edge effect issues.

In mechanical pixel arrays with large pixel aspect ratio, extracting the scintillation light is challenging, and therefore a lower energy resolution is typically observed in these detectors compared to monolithic crystals. We here simulated the light collection efficiency of laser-processed crystals and compared them to monolithic and pixelated detectors (see table 1). It is apparent that in a mechanical array the number of light reflections is larger compared to a monolithic detector, which may give rise to light loss. The two major light loss types are bulk absorption and those related to imperfect surface reflections. In general, light losses due to bulk absorption are more severe in monolithic detectors, which is mainly due to longer traveling distance of individual optical photons. On the other hand, in mechanical arrays there are more light losses due to imperfect surface reflections, especially in arrays with small pixel cross-section and large pixel thickness. Results presented in table 1 demonstrate that regardless of barrier-crystal interface and barrier RI we can collect more light in laser-processed detectors compared to the two other detector categories. This higher light collection efficiency may lead to improved energy resolution as well as improved timing resolution and positioning accuracy. However, a more in depth study is required to methodically study these improvements as the barrier properties investigated in this manuscript are simplified.

We have presented laser-processed LYSO:Ce detectors as a new category between the two mainstreams, being a monolithic crystal and a mechanically pixelated array. We have also shown that by manipulating the barrier RI and the roughness of the barrier-crystal interface, as two out of many barrier parameters, one can achieve both transverse and DOI resolution simultaneously with a single side readout detector scheme. It is apparent that with a huge parameter space one can optimize the optical barrier pattern beyond the simplistic all-the-way and half-way barrier patterns presented here, to fine tune the detector performance characteristics for a specific application.

**Acknowledgements**

LB acknowledges financial support from the Swedish Research Council (VR). This work was supported in part by NIH grant 1R21EB020162-01A1. The authors would like to thank Prof. Mattias Klintenberg at Uppsala University for access to computational resources.